%% file: yb.tex
\documentstyle[12pt,epsfig]{elsart}
\def\Journal#1#2#3#4{{#1} {\bf #2}, #3 (#4)}

\def\PLB{{\em Phys. Lett.}  {\bf B}}
\def\PRL{\em Phys. Rev. Lett.}
\def\PRD{{\em Phys. Rev.} {\bf D}}

\def\PRP{{\em Phys. Rep. }}
\def\EPC{{\em Eur. Phys. J.} {\bf C}}

\def\ZPA{{\em Z. Phys.} A}

\def\PNPP{\em Prog. Nucl. Part. Phys.}
\def\PTP{\em Prog. Theo. Phys.}
\def\PTPS{\em Prog. Theo. Phys. Suppl.}

\def\APP{\em Astroparticle Physics}
\def\EPL{\em Europhys. Lett.}

\def\ra{\rightarrow}
\def\be{\begin{equation}}
\def\ee{\end{equation}}

\newcommand{\ls}{\mbox{$\stackrel{<}{\sim}$ }}
\newcommand{\expe}{experiment}

\newcommand{\osz}{oscillations}

\newcommand{\bb}{double beta decay}
\newcommand{\obb}{0\mbox{$\nu\beta\beta$ - decay}}
\newcommand{\zbb}{2\mbox{$\nu\beta\beta$ - decay}}
\newcommand{\nbb}{neutrinoless double beta decay}
\newcommand{\majo}{Majorana}

\newcommand{\bmbm}{\mbox{$\beta^-\beta^-$} }
\newcommand{\bpbp}{\mbox{$\beta^+\beta^+$} }
\newcommand{\ecec}{\mbox{$EC/EC$} }
\newcommand{\bec}{\mbox{$\beta^+/EC$} }

\newcommand{\bnel}{\mbox{$\bar{\nu}_e$} }

\newcommand{\nel}{\mbox{$\nu_e$}}

\newcommand{\ntau}{\mbox{$\nu_\tau$}}

\newcommand{\ton}{\mbox{$T_{1/2}^{0\nu}$} }

\newcommand{\neu}{neutrino}
\newcommand{\neus}{neutrinos}

\newcommand{\ema}{\mbox{$\langle m_{ee} \rangle$ }}

\newcommand{\gess}{\mbox{$^{76}Ge$ }}
\newcommand{\ybhss}{\mbox{$^{176}Yb$ }}
\newcommand{\hfhss}{\mbox{$^{176}Hf$ }}
\newcommand{\ybhas}{\mbox{$^{168}Yb$ }}
\newcommand{\erhas}{\mbox{$^{168}Er$ }}

\newcommand{\ba}{\begin{array}{c}}
\newcommand{\baz}{\begin{array}{cc}}
\newcommand{\bad}{\begin{array}{ccc}}
\newcommand{\bea}{\begin{equation} \begin{array}{c}}
\newcommand{\eea}{ \end{array} \end{equation}}
\newcommand{\ea}{\end{array}}

\begin{document}
\input front.tex
\input yb1.tex

\input yb2.tex
\input yb3.tex
\input yb4.tex

\input refyb.tex
\newpage
\input pictsyb.tex

\end{document}

%% file: front.tex
\begin{frontmatter}
\title{Double beta decay with large scale Yb-loaded scintillators}
\author{K. Zuber}
\address{Lehrstuhl f\"ur Experimentelle Physik IV, Universit\"at Dortmund,
Otto-Hahn Str.4, 44221 Dortmund, Germany}
\begin{abstract}
The potential of large scale Yb-loaded liquid scintillators as proposed for
solar neutrino spectroscopy are investigated with respect to double beta decay.
The potential for \bmbm{}- decay of \ybhss as well as the \bec{}- decay 
for \ybhas is discussed.
Not only getting for the first time an experimental half-life limit
on \ybhss{}- decay, this
will even
be at least comparable or better than existing ones from different isotopes,
for the first time a realistic chance to detect \bec{}- decay exists.
Effects of MeV-neutrinos are discussed as well.
\end{abstract}
\end{frontmatter}

%% file: yb1.tex
\section{Introduction}
The fundamental question whether \neus{} have a non-vanishing rest mass
is still one of the big open problems of current particle physics.
In case of massive neutrinos a variety of new physical processes open up
\cite{zub98}.
Over the last years evidence has grown for a non-vanishing mass by investigating
solar and atmospheric neutrinos as well as the results coming from
the
LSND
- \expe{}.
They all can be explained within the framework of \neu{} \osz{}
\cite{bil99}.
However oscillations only depend on the differences of squared masses and are therefore
no absolute mass measurements.
Besides that, the question concerning the fundamental character of \neus{}, whether
being Dirac- or Majorana particles, is still unsolved.
A process contributing information to both questions is given by \nbb{} of a nucleus
(Z,A) 
\be
(Z,A) \ra (Z+2,A) + 2 e^-  \quad (\obb)
\ee
This process is violating lepton - number by two units and only allowed if
\neus{} are massive \majo{} particles.
The quantity which can be extracted out of this is called effective \majo{} \neu{} mass
\ema{} and given by
\be
\label{eq:ema}\ema = \mid \sum_i U_{ei}^2 \eta_i m_i \mid
\ee
with the relative CP-phases $\eta_i = \pm 1$, $U_{ei}$ as the mixing
matrix elements and
$m_i$ as the
corresponding mass eigenvalues.
Currently the best limit is given by investigating \gess{} resulting in an
upper bound of \ema $\ls$ 0.2 eV \cite{bau99}. Furthermore to check the
reliability of the calculated
nuclear matrix elements, the standard model process
\be
(Z,A) \ra (Z+2,A) + 2 e^- + 2 \bnel \quad (\zbb)
\ee
is investigated as well.\\
In a more general scheme \obb{} can be mediated by massive Majorana \neus{} or
right-handed currents (also other physics beyond the standard model might be involved,
but this is not discussed here).
To obtain more information on this, it is worthwile
to look for
transitions into excited states, which are dominated by right-handed
currents \cite{doi85}.
Another interesting alternative within this context consists in an
investigation of \bpbp{}-decay \cite{hir94}.
Three different decay channels can be considered here
\bea
(Z,A) \ra (Z-2,A) + 2 e^+ + (2 \nel) \quad \mbox{(\bpbp{})}\\
e_B^- + (Z,A) \ra (Z-2,A) + e^+ + (2 \nel) \quad \mbox{(\bec{})}\\ 
2 e_B^- + (Z,A) \ra (Z-2,A) + (2 \nel) + 2 \mbox{X-rays} \quad
\mbox{(\ecec{})}\\
\eea
The first one is the easiest to detect because of the 2 positrons (corresponding to
four 511 keV photons) but on the other hand is largely suppressed by a low
Q-value
and the Coulomb repulsion of the positrons and the atomic nucleus.
The one with the largest Q-value is the \ecec{}- decay, but difficult to
detect
because only X-rays are emitted. On the other hand this decay could go
into
an excited final state, which might allow a better detection
because of the correlated deexcitation photon. Also here \zbb{} and \nbb{} have to be
distinguished. Current half-life
limits are
in the order of
10$^{20}$ yr obtained with $^{106}$Cd and $^{78}$Kr \cite{bel99,gav00}.\\
Recently a new concept for solar neutrino spectroscopy has been proposed based on 
\bb{} candidates \cite{rag97}. This has been turned into a realistic
detector design (LENS) 
which is using 20t of
Yb in form of a liquid scintillator \cite{len99}. 
A coincidence between the population of an excited state by neutrino
capture (and therefore emission of an electron) and the detection of the
deexcitation photon serves as a signal.
In this paper
the implications of this experiment for \bb{} using the two available isotopes, \ybhas{}
and \ybhss{}, are investigated.

%% file: yb2.tex
\section{\bmbm{}- decay of \ybhss }
First we investigate the decay
\be
\ybhss
\ra \hfhss + 2 e^- (+ 2 \bnel)
\ee 
The decay scheme of \ybhss{} is shown
in Fig.1. The Q-value of the
ground
state transition ($0^+ \ra 0^+$) corresponds to 1079 keV, with a natural
abundance of
this isotope of 12.7 \%. Theres exist no experimental number on \bb{} of
\ybhss{} yet. 
Let us consider the \zbb{} first.
The predicted half-life for the \zbb{} ground transition
is of the order 9.8 $\cdot 10^{21}$yr \cite{sta90}, which would correspond
to a signal of
about 1700
events/day. This number has to be corrected for detection efficiencies and
other experimental parameters. Nevertheless it is an enormous rate for such
a kind of experiment.\\
Also a stringent bound on the \obb{} mode can be obtained.
The achievable half-life can be estimated by
\be
\ton = (1.8 \cdot 10^{24} kg^{-1}) \cdot a \cdot \sqrt{\frac{M \cdot t}{\Delta E \cdot B}}
\ee
where a is the isotopical abundance, M the used mass, t the measuring time,
$\Delta E$ the energy resolution at the peak position and B the background
index, typically given in counts/keV/kg/yr.
Assuming a five year measuring time and an energy resolution of 10 \% 
this would result in $\ton > 1.5 \cdot 10^{26} / \sqrt{B}$ yr.
Therefore even an assumed half-life of only $10^{26}$ yr (already at least a factor of 2
higher than any obtained experimental limit yet) would correspond to 60
events/year. Using the matrix elements of \cite{sta90} 
this 
would lead to a bound on \ema
of
\be
\ema \ls 0.1 eV
\ee
Transitions to excited states can only populate a $2^+$ - state at 88.4 keV.
This state has a life-time of only 1.39 ns which is too short
to use the coincidence technique and a possible transition to that state
would just add to the ground state signal. The phase phase reduction is not
too dramatic for this state, 
but $2^+$ - transitions have in general smaller matrix elements
than the ground state transition \cite{doi92}.

%% file: yb3.tex
\section{\bpbp{}- decay of \ybhas }
For \ybhas{} two decay modes are possible:
\bea
e^-_B + \ybhas \ra \erhas + e^+ (+ 2 \nel) \quad (\bec)\\
2 e^-_B + \ybhas \ra \erhas + (2 \nel) + X-Rays \quad (\ecec)
\eea
The decay scheme of \ybhas{} is shown in Fig. 1. Having a
natural abundance of 0.13 \% the \bec (Q-value = 399 keV) and \ecec{} (Q-value
= 1421 keV) decay modes are possible. No experimental information on any of
these
decay modes for \ybhas{} exist. 
In the following only the 2 $\nu$ decay modes are considered.
In a recent theoretical paper the 2 $\nu$ \ecec ground
state
transition
was estimated and a half life prediction of $T_{1/2}^{EC/EC} = 2 \cdot
10^{23}$
yrs was
obtained
\cite{cer99}. This is in the same order as a recently calculated theoretical half-life
for $\alpha$-decay of \ybhas \cite{fuj00}. Using the fact
that in 
other isotopes the estimated \bec{} half-life is typically one order
of magnitude
higher than for \ecec{} \cite{hir94} , this would also imply that we can expect
$T_{1/2}^{\beta^+ /EC} \approx
10^{24}$ yrs for this mode. 
Using this rough
estimate an event rate of about 180 per year for the \bec{} - decay mode
and roughly 10 times more for the \ecec{} - mode can be expected. 
A coincident signal like in the solar neutrino
case might
be possible here using the characteristic radiation from the EC together with the 2
annihilation photons. 
Maybe even
the transition to an excited $2^+$-state
might be used to increase the signal. 
This state has a life-time of 1.9 ns and decays under emission of a 79.8
keV photon. 
This together with the annihilation photons might form a signal depending on the exact timing.
Also for this state phase space suppression is not too strong. 
For the 2 $\nu$ \ecec{}- decay three more states open up
($2_1^+$ at 821 keV, $0_1^+$ at 1217 keV and $2_2^+$ at 1276 keV). 
The deexcitation photons in principle could serve as a signal, but they are strongly phase
space suppressed. The estimated half-life for the $0_1^+$ state at 1217 keV is 
$5.4 \cdot 10^{33}$ yrs \cite{cer99}.

%% file: yb4.tex
\section{Summary and conclusion}
Using large scale Yb-loaded liquid scintillation detectors
for solar neutrino spectroscopy (LENS) would also allow to probe half-lifes
for 
\bmbm decay of \ybhss{} of the order \ton $\approx 10^{26}$ yrs and beyond, corresponding to
neutrino
mass limits of \ema $\approx$ 0.1 eV and below. With respect to nuclear matrix
element
uncertainties this is an independent probe in the \neu{} mass region below 1 eV currently only
touched by \gess{}.
Furthermore, because the current mass limit on \ntau{} is 18.2 MeV \cite{bar98} it is
still not
excluded that a massive eigenvalue in the MeV-region exists. Such a state 
could lead
to an atomic mass dependent effect in \ema{} because the neutrino mass in
the 
propagator can no longer be neglected as discussed in \cite{hal83,zub97}. Therefore
both isotopes (\gess{} and \ybhss{}) combined provide important informations and would exclude a
mixing between a light
and
heavy state as shown 
Fig. 2.\\
The presence of a significant amount of \ybhas gives for the first time a
realistic chance to observe \bec{} or \ecec{} - decay well within the
range
of theoretical predictions. The proposed solar neutrino experiment LENS
therefore should be designed
in a way to account for the physics topics discussed here as well.
\section*{Acknowledgements}
I would like to thank R. S. Raghavan for useful discussions.

%% file: pictsyb.tex
\begin{center}
\begin{figure}
\begin{tabular}{cc}
\mbox{\epsfig{file=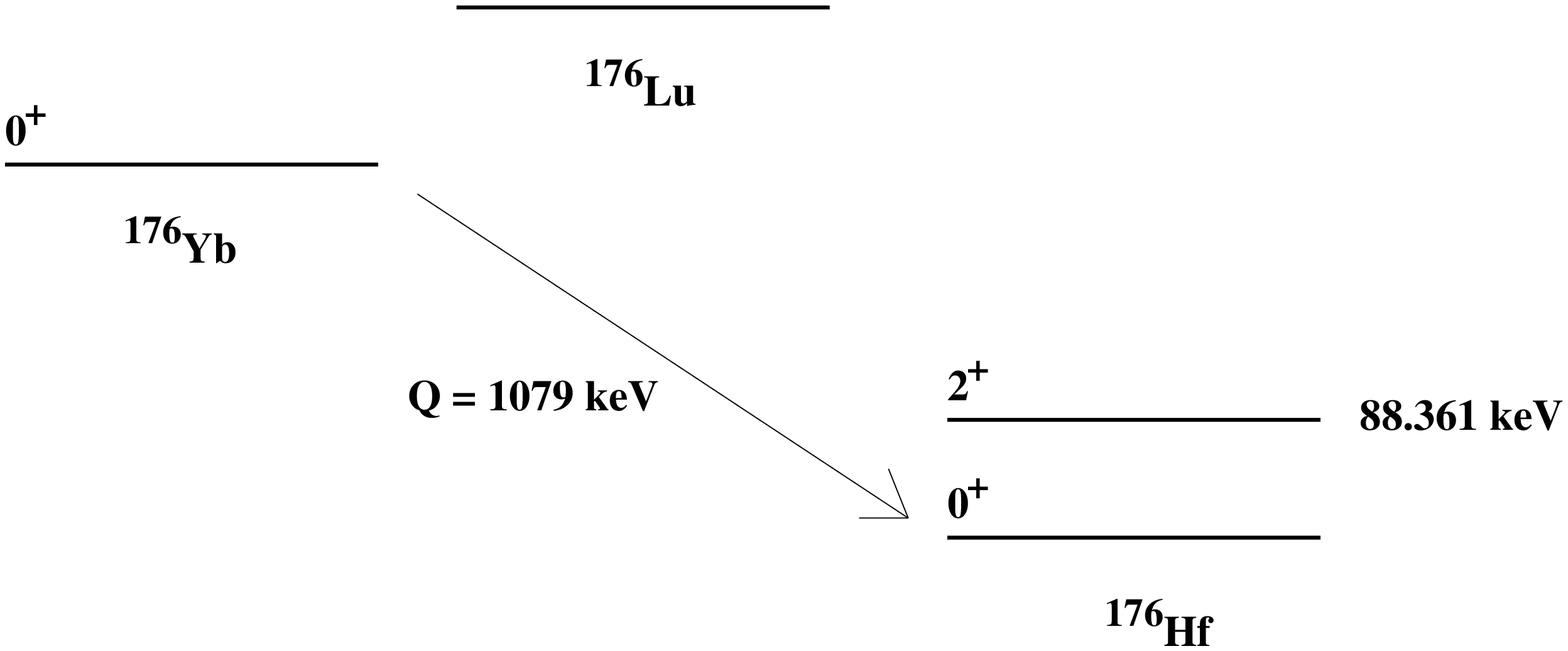,height=4cm,width=7cm}} &
\mbox{\epsfig{file=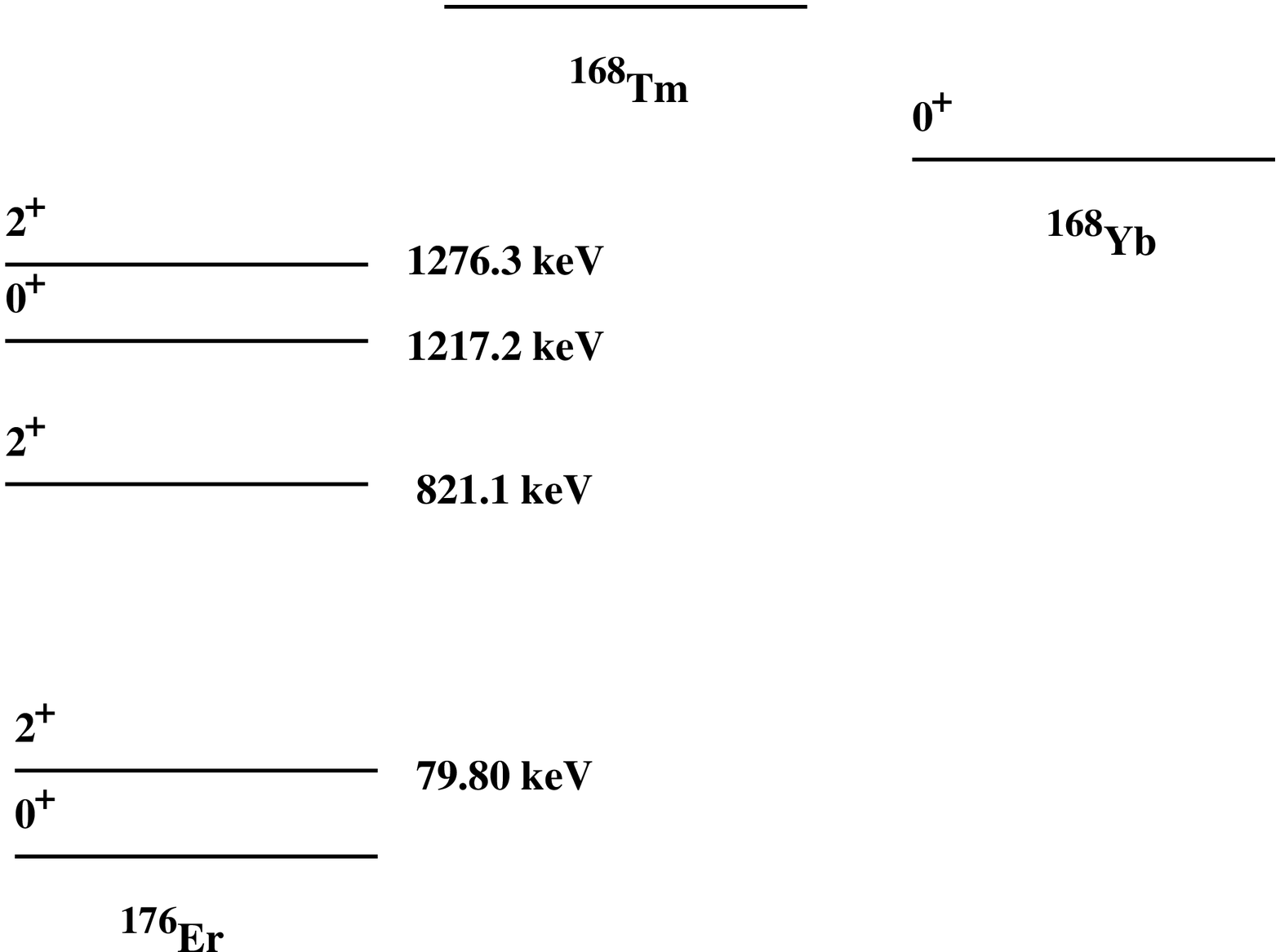,height=5cm,width=7cm}}
\end{tabular}
\vspace{2mm}
\label{pic:levels}
\caption{Nuclear levels which can be populated in \bmbm{} of \ybhss{} (left) and 
\bpbp{} of \ybhas{} (right).
Both isotopes have a low lying $2^+$ - state at about 80 keV. The three high
lying states in \ybhas{}
are only shown for completeness, they are strongly phase space suppressed.}
\end{figure}
\begin{figure}
\mbox{\epsfig{file=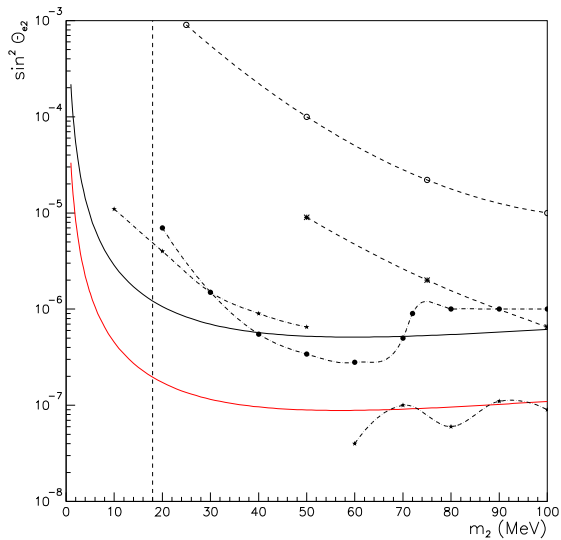,height=8cm,width=8cm}}
\vspace{2mm} 
\label{pic:limits}
\caption{Upper limits on the mixing matrix element $\mid U_{e2}^2 \mid$ of a heavy MeV neutrino
state
$m_2$ to a light one 
due to an A-dependent effect.
Everything on the right to the vertical line is excluded for 
\ntau{} by the ALEPH-results. 
The compilation of the other experimental limits is taken from \protect \cite{zub97}. 
The upper solid curve corresponds to the current status in \bb{} using Ge and Te results. 
Shown as the lower solid curve is the achievable limit
by combining the current \gess{} limit together with an assumed half-life limit on \ybhss{} of $1 \cdot 10^{26}$
yr. Evidently these \bb{} bounds are only valid for \majo{} \neus.}
\end{figure}
\end{center}